\documentclass[pdflatex,sn-standardnature]{sn-jnl}

\usepackage{rotating}
\usepackage{float}
\usepackage{tabularx}
\usepackage{makecell}
\usepackage{lmodern}
\usepackage{anyfontsize}
\usepackage{amsmath}
\usepackage{cleveref}

\begin{document}

\title[Artificial Intelligence for Quantum Computing]{Artificial Intelligence for Quantum Computing}




\author[1]{\fnm{Yuri} \sur{Alexeev\textsuperscript{\dag}}}
\author[1]{\fnm{Marwa H.} \sur{Farag\textsuperscript{\dag}}}
\author[1]{\fnm{Taylor L.} \sur{Patti\textsuperscript{\dag}}}
\author*[1]{\fnm{Mark E.} \sur{Wolf\textsuperscript{\dag}}}\email{mawolf@nvidia.com}
\author[2]{\fnm{Natalia} \sur{Ares}}
\author[3,4]{\fnm{Al\'an} \sur{Aspuru-Guzik}}
\author[5,6]{\fnm{Simon C.} \sur{Benjamin}}
\author[5,6]{\fnm{Zhenyu} \sur{Cai}}
\author[1]{\fnm{Shuxiang} \sur{Cao}}
\author[1]{\fnm{Christopher} \sur{Chamberland}}
\author[1]{\fnm{Zohim} \sur{Chandani}}
\author[2]{\fnm{Federico} \sur{Fedele}}
\author[1]{\fnm{Ikko} \sur{Hamamura}}
\author[1]{\fnm{Nicholas} \sur{Harrigan}}
\author[1]{\fnm{Jin-Sung} \sur{Kim}}
\author[1]{\fnm{Elica} \sur{Kyoseva}}
\author[1]{\fnm{Justin G.} \sur{Lietz}}
\author[1]{\fnm{Tom} \sur{Lubowe}}
\author[1]{\fnm{Alexander} \sur{McCaskey}}
\author[7,8]{\fnm{Roger G.} \sur{Melko}}
\author[1]{\fnm{Kouhei} \sur{Nakaji}}
\author[9]{\fnm{Alberto} \sur{Peruzzo}}
\author[1]{\fnm{Pooja} \sur{Rao}}
\author[1]{\fnm{Bruno} \sur{Schmitt}}
\author[1]{\fnm{Sam} \sur{Stanwyck}}
\author[10]{\fnm{Norm M.} \sur{Tubman}}
\author[11]{\fnm{Hanrui} \sur{Wang\textsuperscript{11}}}
\author[1]{\fnm{Timothy} \sur{Costa}}

\affil[1]{\orgname{NVIDIA Corporation}, \orgaddress{\street{2788 San Tomas Expressway}, \city{Santa Clara}, \postcode{95051}, \state{CA}, \country{USA}}}

\affil[2]{\orgname{Department of Engineering Science, University of Oxford}, \orgaddress{\street{Parks Road}, \city{Oxford}, \postcode{OX1 3PJ}, \country{United Kingdom}}}

\affil[3]{\orgname{Department of Chemistry, Computer Science, Materials Science and Engineering, and Chemical Engineering and Applied Science,University of Toronto}, \orgaddress{\street{80 St George St}, \city{Toronto}, \postcode{M5S 3H6}, \state{ON}, \country{Canada}}}

\affil[4]{\orgname{Vector Institute for Artificial Intelligence}, \orgaddress{\street{661 University Ave Suite 710}, \city{Toronto}, \postcode{M5G 1M1}, \state{ON}, \country{Canada}}}

\affil[5]{\orgname{Quantum Motion \footnote{\dag These authors contributed equally to this work.}}, \orgaddress{\street{9 Sterling Way}, \city{London}, \postcode{N7 9HJ}, \country{United Kingdom}}}

\affil[6]{\orgname{Department of Materials, University of Oxford}, \orgaddress{\street{Parks Road}, \city{Oxford}, \postcode{OX1 3PH}, \country{United Kingdom}}}
\affil[7]{\orgname{Department of Physics and Astronomy, University of Waterloo}, \orgaddress{\street{200 University Avenue West.}, \city{Waterloo}, \postcode{N2L 3G1}, \state{ON}, \country{Canada}}}

\affil[8]{\orgname{Perimeter Institute for Theoretical Physics}, \orgaddress{\street{ 31 Caroline Street North}, \city{Waterloo}, \postcode{N2L 2Y5}, \state{ON}, \country{Canada}}}

\affil[9]{\orgname{Qubit Pharmaceuticals}, \orgaddress{\street{29, rue du Faubourg Saint Jacques}, \city{Paris}, \postcode{75014}, \country{France}}}

\affil[10]{\orgname{NASA Ames Research Center}, \orgaddress{\street{Moffett Field}, \city{California}, \postcode{94035-1000}, \country{USA}}}

\affil[11]{\orgname{UCLA Computer Science Department}, \orgaddress{\street{404 Westwood Plaza, Engineering VI}, \city{Los Angeles}, \postcode{90095-1596}, \state{CA}, \country{USA}}}

\abstract{Artificial intelligence (AI) advancements over the past few years have had an unprecedented and revolutionary impact across everyday application areas. Its significance also extends to technical challenges within science and engineering, including the nascent field of quantum computing (QC). The counterintuitive nature and high-dimensional mathematics of QC make it a prime candidate for AI's data-driven learning capabilities, and in fact, many of QC's biggest scaling challenges may ultimately rest on developments in AI. However, bringing leading techniques from AI to QC requires drawing on disparate expertise from arguably two of the most advanced and esoteric areas of computer science. Here we aim to encourage this cross-pollination by reviewing how state-of-the-art AI techniques are already advancing challenges across the hardware and software stack needed to develop useful QC - from device design to applications. We then close by examining its future opportunities and obstacles in this space.}


\maketitle


\tableofcontents

\section{Introduction}\label{sec:intro}
Quantum computing (QC) has the potential to impact every domain of science and industry, but it has become increasingly clear that delivering on this promise rests on tightly integrating fault-tolerant quantum hardware with accelerated supercomputers to build accelerated quantum supercomputers. Such large-scale quantum supercomputers form a heterogeneous architecture with the ability to solve certain otherwise intractable problems. Many of these problems, such as chemical simulation or optimization, are projected to have significant scientific, economic and societal impact~\cite{alexeev2021quantum}. 

However, transitioning hardware from noisy intermediate-scale quantum (NISQ) devices to fault-tolerant quantum computing (FTQC) faces a number of challenges. Though recent quantum error correction (QEC) demonstrations have been performed\cite{bluvstein2024logical, ryan2021realization}, all popular qubit modalities suffer from hardware noise preventing the below-threshold operation needed to perform fault-tolerant computations. But, even qubits performing below threshold face scaling obstacles. FTQC is demanding, and necessitates more resourceful QEC codes, faster decoder algorithms and carefully designed qubit architectures. Both QC hardware research and current quantum algorithms also require further development, with explorations of more resource-efficient techniques having the potential to dramatically shorten the road-map to useful quantum applications.

Though high performance computing (HPC) \cite{li2024quantum, arute2019quantum, erdman2022model}, and in particular, accelerated GPU computing \cite{willsch2022gpu, thomson2024unravelling}, already drives QC research through circuit and hardware simulations, the rise of generative artificial intelligence (AI) paradigms \cite{bandi2023power} has only just begun. Foundational AI models \cite{zhou2023comprehensive}, characterized by their broad training data and ability to adapt to a wide array of applications, are emerging as an extremely effective way to leverage accelerated computing for QC. While the architecture landscape of these models is diverse,  transformer models \cite{vaswani2017attention} have proven particularly powerful, and especially popularized by OpenAI's generative pre-trained transformer (GPT) models \cite{achiam2023gpt,yenduri2024gpt}. There is already a strong precedent for these models being applied to technical yet pragmatic tasks in other fields, ranging from biomedical engineering \cite{cheng2023exploring} to materials science \cite{liu2023generative}. Bringing the deep utility and broad applicability of such models to bear on the problems facing QC is a key goal of this review.

    \begin{figure}[h]
        \centering
        \includegraphics[width=5in]{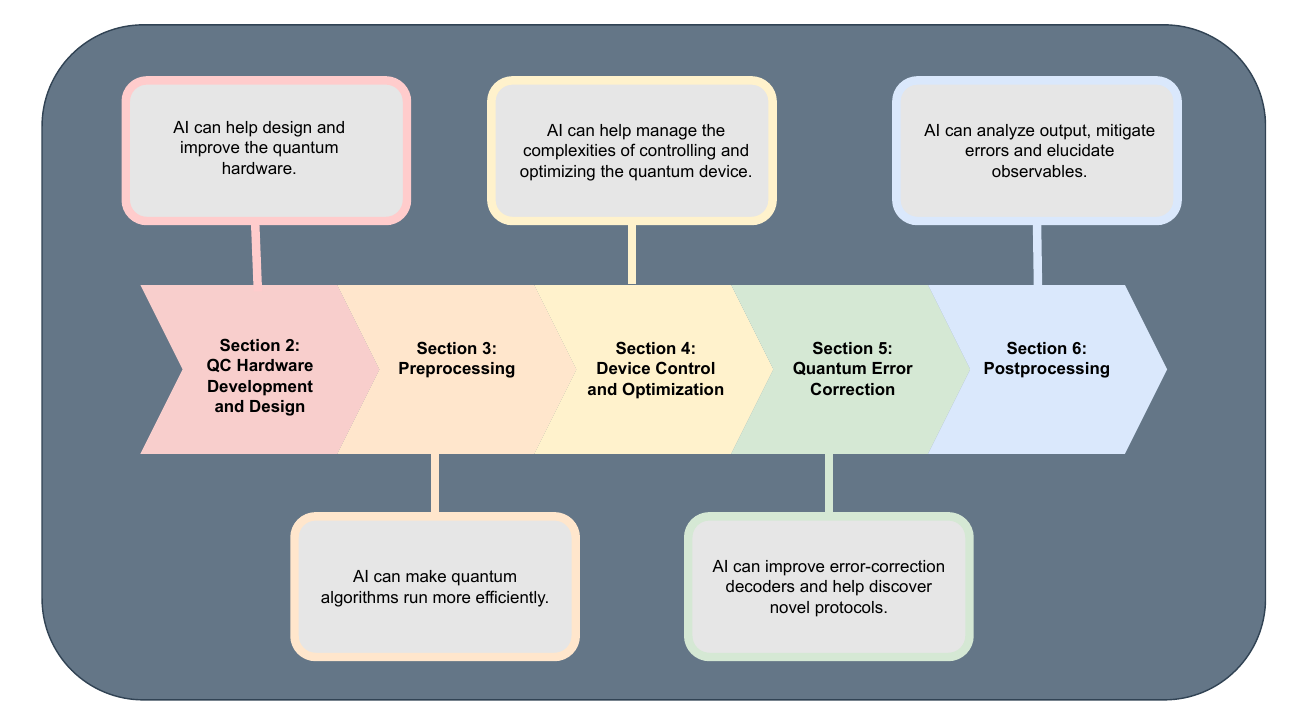}
        \caption{A depiction of the sections covered in this review and how AI can be used to benefit the entire QC stack.}
        \label{fig:TOC}
        \end{figure}

There is ample intuition to motivate exploring AI as a breakthrough tool for QC. The inherent nonlinear complexity of quantum mechanical systems \cite{dunjko2023artificial} makes them well-suited to the high-dimensional pattern recognition capabilities and inherent scalability of existing and emerging AI techniques~\cite{hornik1989multilayer}. Many AI for quantum applications are being realized for near term development of quantum computers and the long-term operation of scalable FTQC workflows. This review examines applications of state-of-the-art AI techniques that are advancing QC with the goal of fostering greater synergy between the two fields. 

It is worth clarifying that this review focuses solely on the impact of AI techniques for developing and operating useful quantum computers (AI for quantum) and does not touch upon the longer-term and more speculative prospect of quantum computers one day enhancing AI techniques (often referred to as quantum for AI), which are surveyed in Ref. \cite{peral2024systematic}.

The content of this review is organized according to the causal sequence of tasks undertaken in operating a quantum computer (Fig. \ref{fig:TOC}). We immediately stretch this taxonomy by beginning in Sec. \ref{sec:dev_design} with how AI techniques can accelerate fundamental research into designing and improving the quantum hardware needed to operate a useful device. Then Secs. \ref{sec:AI_for_Pre}, \ref{sec:control_calibration}, \ref{sec:qec}, and \ref{sec:AI_for_post} step through AI's roles in the widely accepted QC workflow: preprocessing, tuning, control and optimization, QEC, and postprocessing. AI's use in algorithm development, that is, AI's impact on various algorithmic subroutines, is covered throughout the whole manuscript, where relevant, and spans many tasks across the workflow. The review concludes with Sec. \ref{sec:outlook} by looking ahead to fruitful areas where AI might still be applied and speculating on areas of development that will further AI's ability to solve QC's remaining challenges.

\subsection{A Brief Survey of AI Methods}\label{sec:AIMethodsSurvey}

\begin{figure}[h]
        \centering
        \includegraphics[width=\textwidth]{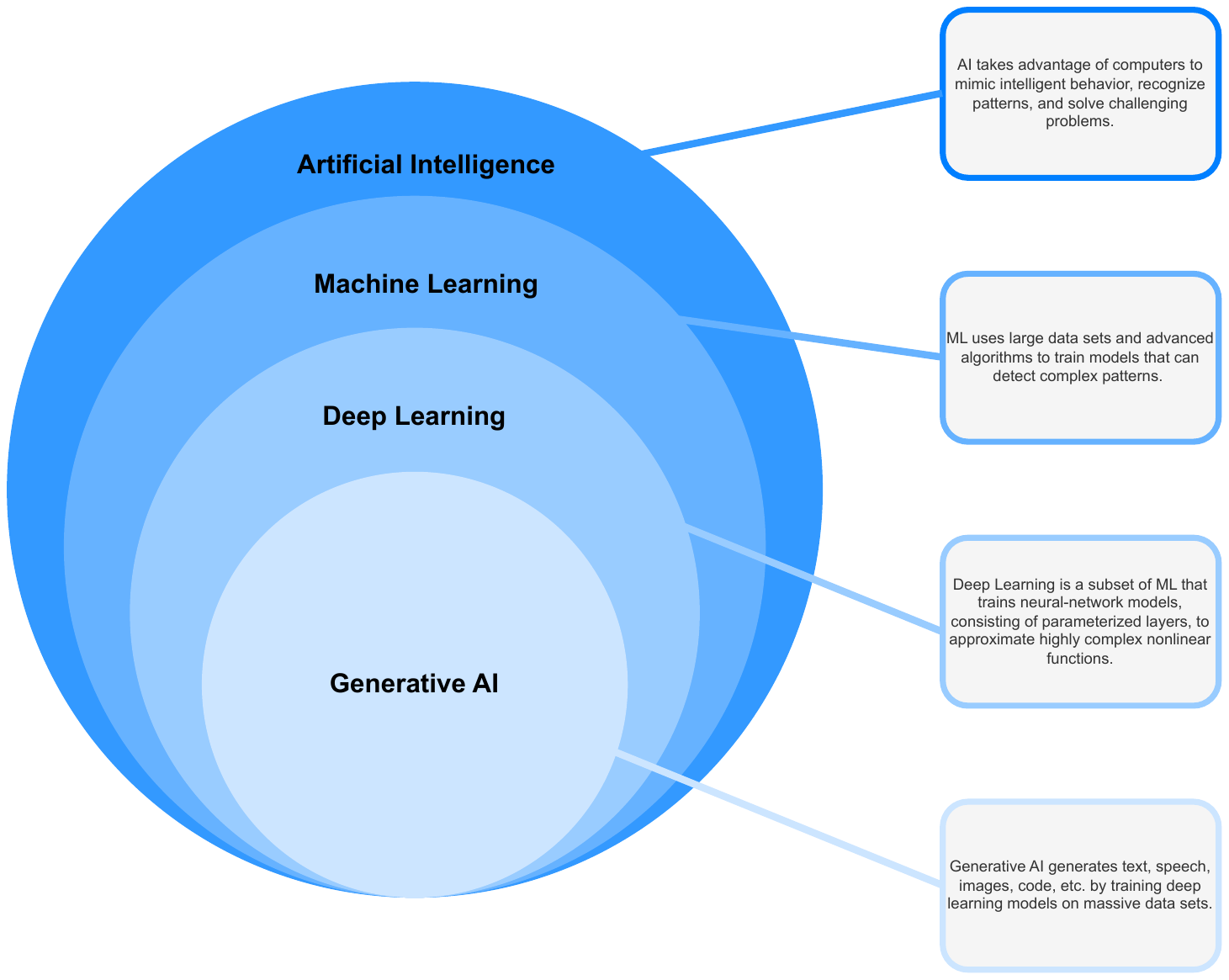}
        \caption{A simple hierarchy from Artificial Intelligence to generative AI, broadly contextualizing the techniques discussed in this work. Each level is paired with a simple description.}
        \label{fig:aichart}
        \end{figure}

The majority of modern AI methods pertain to the subfield of machine learning (ML)\footnote{AI is a strict superset of ML, since there are AI algorithms based on hard-coded rules, which there is no training procedure that algorithmically 'learns' from data.}, consisting of algorithms that extract and utilize information from datasets \cite{zhuhadar2023application}. Though there are many different ML architectures (decision trees, support vector machines, clustering models, etc.), in this review we focus on architectures pertaining to the field of ``deep learning'', meaning that they are based on some form of deep neural network (DNN) \cite{lecun2015deep,janiesch2021machine}. Maturing to industrial scales in the 2010's, DNNs learn multiple data abstractions via the process of backpropagation. These data abstractions are used to construct useful representations of the dataset of interest. DNNs are characterized by their flexibility in representing patterns in data and the adaptability of their architectures. This has resulted in DNNs contributing numerous architectures to the sprawling phylogeny of ML models, which have found application across disciplines. We depict these relationships in Fig. \ref{fig:aichart}.

In the broadest of strokes, we can categorize DNN applications as discriminative and generative \cite{bernardo2007generative}. The former seeks to learn the conditional probability distribution $P(\mathbf{y} {\mid} \mathbf{x})$ of value vector $\mathbf{y}$ given feature vector $\mathbf{x}$, whereas the latter seeks the joint probability distribution $P(\mathbf{x},\mathbf{y})$. Less formally, discriminative models learn to distinguish between data types while generative models learn to produce new instances of their target data classes.

DNN architectures are incredibly diverse and ever expanding, and we here overview but a handful of them that will prove germane in this review. Among the most popular of such architectures is Reinforcement Learning (RL), which focuses on sequential decision-making \cite{arulkumaran2017deep,shakya2023reinforcement}. In RL, the DNN or ``agent'' is tasked with navigating a learning problem and trained by assigning it a score after each decision it makes, rewarding it for useful decisions and punishing it for problematic ones. In maximizing this so-called cumulative ``reward'', the agent learns how to evaluate outcomes (value) and respond to situations (policy). RL models find particular utility for automating multi-faceted tasks in dynamic environments with delayed rewards. However, RL is often difficult to implement due to its sensitivity in selecting hyperparameters.

ML tasks often focus on learning from and producing new sequences. A canonical example is natural language processing, wherein the ML agent learns from existing sentences (word sequences) to produce new ones \cite{chowdhary2020natural,khurana2023natural}. For many years, such problems were addressed with recurrent neural networks (RNNs), which apply a single set of weights along the elements of a given input sequence, producing an output sequence step-by-step \cite{irsoy2014deep,socher2011parsing}. More recently, Transformer models \cite{vaswani2017attention,han2022survey}, including the famed GPT family \cite{achiam2023gpt}, have dominated sequence learning, bolstered by their parallelizability, long-range and bi-directional token (word) context, and their adroitness with variable-length inputs.

Another popular model for generative tasks is the diffusion model \cite{ho2020denoising}, such as more recent DALL-E models \cite{ramesh2022hierarchical}. These models use random walks with drift, as formalized by, e.g., Markov chains, to gradually add noise to target data and then learn the reverse denoising process. After such training is complete, the diffusion models can generate desired samples from noise.

Critical for training all of these deep learning methods is high-quality data.  In the case of QC, this data must often be obtained via simulation with supercomputers due to noise and scale limitations of quantum computers, as well as the cost (time and economic) of obtaining quantum data. Section \ref{sec:simulation} discusses simulation in greater detail.

\section{AI for Quantum Computer Development and Design}\label{sec:dev_design}

Fundamental improvements to quantum hardware currently requires precise, costly and extremely difficult experimentation. From design to fabrication, characterization and control, AI can accelerate the workflows surrounding the quantum device development cycle - providing insights into the complexities of quantum systems and reducing the timeline for developing quantum computers. In this section we cover three main areas where AI is already being applied to development and design.

\subsection{System Characterization}\label{sec:system_char}
Central to the design of future  useful quantum devices is our understanding of today's smaller experimental quantum systems. Such studies are led by the broad field of Hamiltonian Learning \cite{wiebe2014hamiltonian, GentileLearningModels2021, GebhartLearningQuantumSystems2023}, which seeks to identify the generating Hamiltonian of observed quantum dynamics through the use of ML methods. Such methods are quite generally applicable \cite{flynn2022quantum,sarma2024precision} and applying these to characterize measurement-expensive and noise-prone  contemporary quantum computers \cite{preskill2018quantum} can be challenging. There has been success in meeting these constraints by refining models to require only tractable amounts of quantum input data \cite{che2021learning} and adapting them to be suitable for learning the non-Markovian dynamics relevant to noisy systems \cite{banchi2018modelling}. While challenging, the ML-assisted characterization of quantum systems can be greatly simplified by the inclusion of relevant information~\cite{gebhart2023learning}, e.g., observable constraints, as has proven useful in so-called ``graybox'' models \cite{youssry2020modeling, krastanov2020unboxing}, which combine ``whitebox'' physics equations to guide ``blackbox'' deep learning models.

       \begin{figure}[h]
        \centering
        \includegraphics[width=\textwidth]{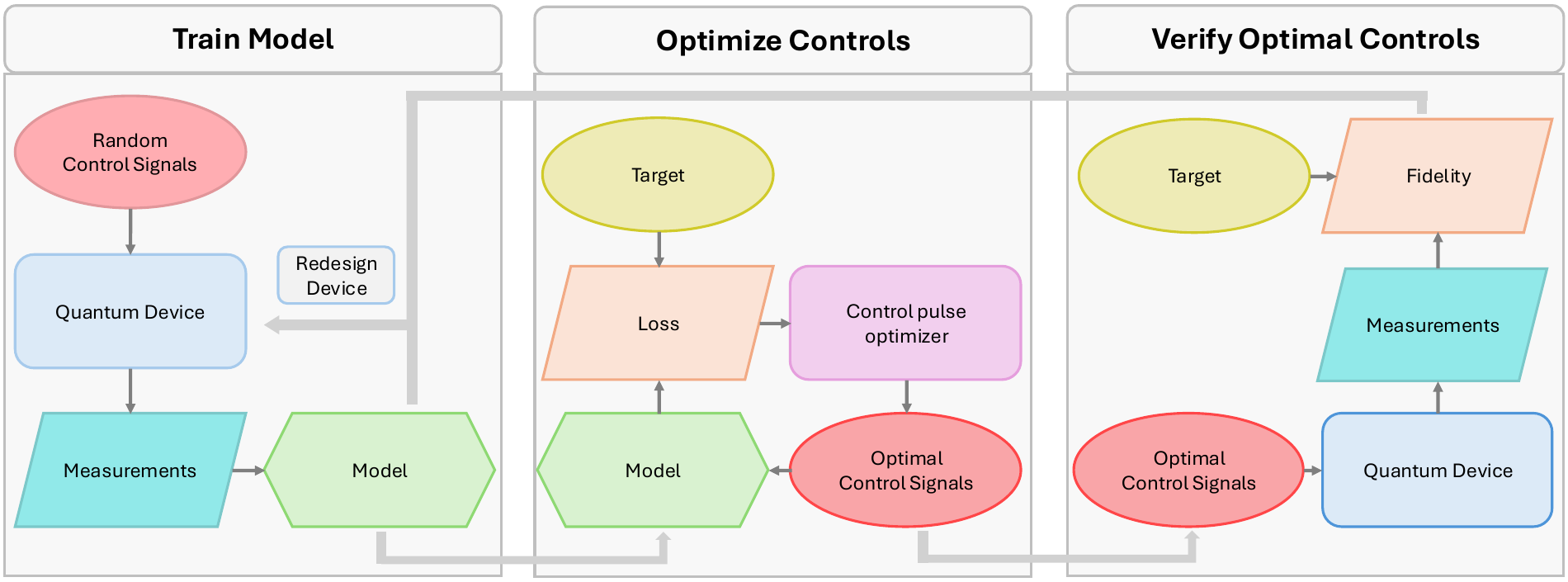}
        \caption{Schematic of process for training ML models, adapted from~\cite{youssry2024experimental}.
        The process starts with creating an experimental or simulated dataset by applying controls to the system and recording outputs. This dataset trains ML models, which are then use to determine optimal control settings.
        }
        \label{fig:photonic_pulse_opt}
        \end{figure}

Graybox models also have utility in more applied tasks, such as optimizing control voltages for a photonic quantum circuits (Fig. \ref{fig:photonic_pulse_opt}) \cite{youssry2024experimental}. More generally, ML methods have been used to learn quantum device characteristics otherwise inaccessible to experiments - such as disorder potentials \cite{craig2024bridging, percebois2023} and the nuclear environment of a qubit~\cite{jung2021deep}. ML-based algorithms can also be used to tune, characterize and optimize qubit operation (see section \ref{sec:control_calibration}), take data efficiently~\cite{lennon2019efficiently} and learn the Hamiltonian parameters of a variety of quantum systems~\cite{gebhart2023learning}. 

\subsection{Platform Design}\label{sec:platform_design}
The design and manufacture of QC devices requires not only physically characterizing them at the materials level, but also analyzing the individual components they're used in, which can vary widely due to inevitable irregularities (e.g., in fabrication \cite{siddiqi2021engineering}), nonidealities (e.g., crystal strain \cite{marshall2022high}), and the limitations of optical components \cite{scully1997quantum}. AI approaches have been employed to successfully design multi-qubit operations by exploring potential superconducting designs that were then demonstrated experimentally \cite{Menke2021-yz,Menke2022-qc}.  In addition, AI can be employed to design quantum optical setups that can then be employed to generate highly entangled states \cite{Krenn2021-jq,Flam-Shepherd2022-ri,Cervera-Lierta2022-on}.

AI models learning how to optimize the performance of multi-qubit gates in nonuniform semiconductor-based qubits can automate the handling of manufacturing variabilities in these devices \cite{severin2021cross}. Similarly, AI-automated qubit initialization protocols can be used to navigate the otherwise unintuitive process of preparing fault-tolerant operations at relatively high temperatures using Markov models \cite{huang2023high}, suppressing system temperatures with RL \cite{sarma2022accelerated}, or maximizing the coherence of quantum state transfer with RL \cite{porotti2019coherent}. A distinct yet related challenge is designing AI protocols which are dependent on information that cannot be directly included in model training, such as unavoidable classical noise or inherent quantum uncertainty \cite{fouad2024model}. Such limitations have also been addressed by iterative applications of transfer learning \cite{zhuang2020comprehensive}, e.g., by pre-optimizing system control in progressively more realistic and challenging substrates \cite{van2024cross}. The reverse tactic has likewise proven fruitful, where AI models versed in well-understood quantum systems are used to propose novel quantum experiments. This same technique can be applied to leverage AI models trained on a quantum system to extrapolate new architectures and quantum information protocols \cite{krenn2016automated}.

\subsection{Gate and Pulse Optimization}\label{sec:gate_pulse_opt}
Deep learning methods, particularly RL techniques \cite{shakya2023reinforcement}, have proven especially fruitful across qubit modalities for one particular aspect of implementing logical operations - pulse optimization. Whilst AI approaches to pulse engineering have been successfully applied to more native operations, such as scheduling near-adiabatic transitions \cite{ding2021breaking}, it has been most extensively explored for facilitating quantum logic gates in the traditional QC paradigm \cite{koutromanos2024control}. Of particular interest is exploring optimal pulse sequences for qubits on specific quantum hardware, such as superconducting transmon \cite{nguyen2024reinforcement} and charge qubits \cite{giounanlis2021cmos}, as well as quantum dots \cite{daraeizadeh2020designing}. Such studies often target specific limitations of modern devices, such as robustness against quantum noise and environmental distortion \cite{daraeizadeh2020machine}, or decreasing the clock speed of quantum operations while counterbalancing state leakage \cite{wright2023fast}.

\section{AI for Preprocessing}\label{sec:AI_for_Pre}
\label{sec:preprocessing}

Preparing quantum algorithms to run on a quantum device is a significant challenge. Practical implementation of algorithms requires generating compact circuits that run as fast and efficiently as possible, whilst accounting for device-specific constraints. We refer to this process as ``preprocessing". Recent advancements in AI methods have opened new possibilities for more efficient and flexible quantum circuit design. This section surveys how AI is being used to improve key preprocessing steps such as unitary synthesis, circuit reduction, and state preparation - unlocking more efficient quantum algorithms.

\subsection{Quantum Circuit Synthesis}\label{sec:circuit_synthesis}

Circuit synthesis is the orchestration of potentially hardware-specific operations to efficiently realize some desired quantum circuit. Optimizing circuit synthesis is a challenging task that has specific bottlenecks due to the nature of optimizing on quantum hardware, and its complexity quickly becomes unmanageable for larger circuits. In this section, we review several approaches that leverage AI to automate circuit synthesis and overcome its challenges~\cite{kulshrestha2023qarchsearch, barrenplateaus,allen2019convergence,wang2021noise,anschuetz2022quantum}.

\subsubsection{Unitary Synthesis}\label{sec:unitary_synthesis}

Unitary synthesis is a particularly important circuit synthesis task that prepares a quantum circuit to implement a specific unitary operation. The primary challenge is decomposing the unitary matrix representing the operation into a sequence of elementary quantum gates, typically from a universal gate set
\cite{shende2006}.  The complexity of unitary synthesis increases exponentially with the number of qubits, making exact synthesis computationally prohibitive for large quantum systems. Managing this high dimensionality alongside hardware constraints (such as variable gate fidelities and qubit connectivity) necessitates the use of approximation techniques or heuristic methods.

In recent years, AI-based methods have emerged as powerful tools for unitary synthesis. Deep learning techniques can automate navigating the vast space of potential gate sequences during the decomposition process \cite{bukov2018}. For example, RL can treat synthesis as a sequential decision-making problem, where an agent iteratively selects gates to construct a circuit that closely approximates the desired unitary operation.

Moreover, diffusion models \cite{ho2020denoising} - generally considered to be the precursors to transformers, have also recently been applied to generate valid circuits for arbitrary unitary operations  \cite{F_rrutter_2024}. A circuit representation is tokenized such that it can be embedded as a 3-D tensor. A conditioning step follows which ensures that a text prompt is embedded via a pre-trained language model. This along with the unitary representation of the circuit is fed into a convolutional neural network (CNN) architecture known as a U-Net \cite{unet}, which follows the typical diffusion model training procedure. For a given epoch, a timestep, $t$, is sampled and Gaussian noise added to the input data based on $t$. The job of the NN is to learn the added noise via backpropagation. The trained model can then be used during inference to generate valid data distributions from noisy samples. Results in \cite{F_rrutter_2024} demonstrate applications to 3 qubit models with a gate set comprising of 12 gates.

\subsubsection{AI Models to Generate Compact Circuits}\label{sec:compact_circuits}
    

An important requirement for preprocessing is to generate \textit{compact} quantum circuits. Compared to previously considered `brute-force' approaches to quantum circuit generation \cite{PeruzzoVQE2014}, both generative and RL AI models have demonstrated promise in generating more compact circuits \cite{ruiz2024quantum, GPT-QE}. Google DeepMind's RL-based approach to circuit generation, AlphaTensor-Quantum, is a noteworthy example \cite{ruiz2024quantum}. The approach optimizes circuits by minimizing the count of notoriously expensive non-Clifford T-gates - translating the optimization problem into a tensor decomposition. To mitigate challenges associated with RL such as exploring a large combinatorial space, training instability, and computational overhead, domain-specific knowledge is utilized, including alternative implementations of T-gates with Toffoli and controlled-S gadgets.

\begin{figure}[h]
\centering
\includegraphics[width=\textwidth]{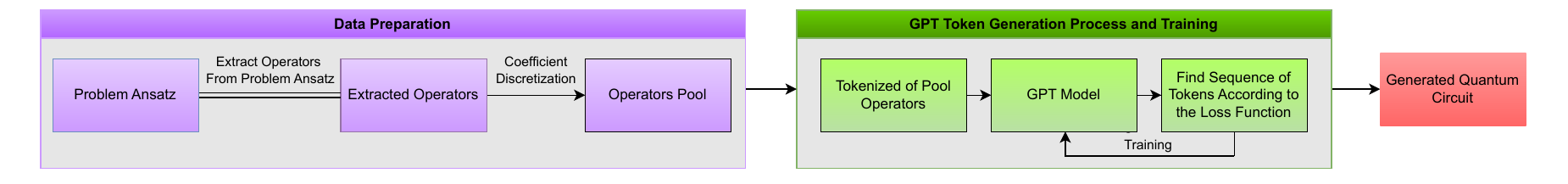}
\caption{Workflow of the GPT-QE algorithm. During the initial \textbf{Operator Preparation} stage, operators are extracted (the choice of operators depends on the problem ansatz - UCCSD and QAOA being two examples), resulting in Hermitian operators $\{P_j\}_{j}$ such as Pauli strings. Additionally, a range of discrete coefficients ($\{\theta_k\}_k$) are generated. $\{P_j\}$ and $\{\theta_k\}$ are combined into different unitary pool operators (\{e$^{i P_j \theta_k}\}_{j,k}$). During the next \textbf{GPT token-generation and training} stage, the \{e$^{i P_j \theta_k}\}$ are tokenized and  passed to a transformer for training. In training, the model produces sequences of tokens for which the loss function is computed. These losses are used to update the transformer parameters. Finally, after training, the model is able to generate a `prediction' of a quantum circuit.}
        \label{fig:gqe}
        \end{figure}

Generative AI has also demonstrated great utility for generating compact circuits. Recent work introduced a generative pre-trained transformer based quantum eigensolver (GPT-QE) depicted in Fig. \ref{fig:gqe}~\cite{GPT-QE}. Here a (GPT) model is employed to sample a quantum circuit sequence from a pre-defined pool of operators (such as UCCSD). The transformer model parameters and energies computed from the sampled quantum circuits are used to compute a loss function. Model parameters are then updated by back-propagation. Repeating this process trains the transformer model to generate quantum circuits minimizing the loss function. GPT-QE models trained for problems in one domain using can be used to construct datasets for pre-training GPT model applicable to adjacent problems. This extensible feature of GPT-QE allows maximum value to be extracted from quantum circuit datasets.


\subsection{Circuit Parameter Learning and Parameter Transfer}\label{sec:param_trans}

Another important strategy that can be considered during the preprocessing stage of quantum computation is whether parameters can be \textit{transferred} between quantum circuits. This is particularly relevant for circuits implementing the Variational Quantum Eigensolver (VQE), the Quantum Approximate Optimization Algorithm (QAOA) and other variational quantum algorithms\cite{Bharti2022-dz,falla2024graph, galda2023similarity,sud2024parameter}. 
Parameter transfer is the process of using optimal circuit parameters from other use cases to accelerate the generation of optimal parameters in a new, distinct use case. 

Graph embedding techniques, such as Graph2Vec~\cite{narayanan2017graph2vec} and GL2Vec~\cite{hong2019gl2vec}, have been used to facilitate such transferability, by identifying structural similarities between graphs representing different problem instances. These embeddings allow AI models to predict optimal parameters for new instances by leveraging pre-optimized donor parameters, significantly reducing the computational overhead compared to running the optimization from scratch~\cite{galda2021transferability}. 

This methodology is especially effective in mitigating the barren plateau problem~\cite{larocca2024review}, wherein a very flat optimization landscape causes gradient-based optimization methods to struggle to navigate toward global optima - a critical issue in the training of quantum circuits. Transferability pipelines built on graph embeddings also allow the scaling of QAOA performance, with an order of magnitude improvement in efficiency under both ideal and noisy conditions~\cite{langfitt2023parameter}.

It is important to mention one particular set of AI algorithms used to improve ML models by using meta-learning.  Meta-learning is a class of algorithms that `learn to learn'.  These algorithms have been successfully applied for circuit initialization ~\cite{2019arXiv190705415V} and circuit optimization~\cite{wilson2021optimizing}.

\subsection{State Preparation}\label{sec:state_prep}

Another preprocessing task prescribed by a number of algorithms is the preparation of particular quantum states. However, naïve implementations of such state preparations generally require circuits having a depth that grows exponentially with problem size \cite{araujo2021divide}. This quickly becomes intractable for large-scale algorithms, motivating a more innovative approach to state preparation. This has led to the exploration of ML-assisted methods \cite{arrazola2019machine} - including both classical and quantum NNs and other related techniques. 

AI-based approaches to state preparation are broad, accommodating the many specialized heuristics, optimizations and initializations that can apply to the wide range of possible state preparation problems~\cite{2023arXiv230613126S}. Many of the techniques already described in this review, such as GPT-QE and meta-learning have been co-opted for state preparation purposes. we note that pre-optimization ideas have also been referred to as ``warm-starting" and ``no-optimization" with the main idea being to use heurstic or classical simulations before starting any optimization on quantum hardware~\cite{PRXQuantum.4.030307,mullinax2023large,khan2023preopt}, which draw on a wide range of ML methods. As quantum algorithms become increasingly refined, optimization tasks are likely to be moved from quantum to classical hardware wherever possible - increasing the relevance of improved AI techniques for state preparation. 

Particular attention has been paid to using RL for state preparation, which has proven particularly successful when employing discrete action spaces \cite{zhang2019does}. RL has been used for state preparation on both ideal \cite{liu2023quantum,Ostaszewski2021Reinforce} and experimental \cite{wang2024arbitrary} systems, and has been used to optimize experimental figures of merit such as fidelity, gate cost and runtime. Remarkably, these approaches have also yielded theoretical insights in addition to blackbox functionality, generalizing to learn entire state classes rather than single instances \cite{haug2020classifying} and adhering to spin glass-like hardness guarantees \cite{bukov2017machine}. 

Finally, we highlight other optimization techniques that have been developed to search over circuit space beyond the techniques described previously in this section.  While there are numerous approaches that have been considered, some of the most promsing techniques includes basin hopping optimization~\cite{burton2022disco},genetic algorithms~\cite{sünkel2023genetic,chivilikhin2020genetic} and Bayesian optimization~\cite{sorourifar2024bayes,Duffield_2023}.  In table \ref{tab:stateprep}, we summarize different state preparation approaches discussed in this section.

\begin{table}[h]
\caption{Circuit synthesis techniques for State Preparation}
\label{tab:stateprep}
\begin{tabular}{p{5cm}p{3cm}p{3cm}}
\hline
Reference  & Method  & Related Refs.\\
\hline
\cite{grimsley2019adaptive}~Grimsley et al. (2019) &  Search(operator pool) &\\

\cite{burton2022disco}~Burton et al. (2022) & Search(Basin Hopping) &\\

\cite{Duffield_2023}~Duffield et al. (2023) & Bayesian Optimization &~\cite{sorourifar2024bayes}\\

\cite{chivilikhin2020genetic}~Chivilikhin et al. (2020) & Genetic Algorithm & ~\cite{sünkel2023genetic} \\ 
\cite{GPT-QE}~Nakaji et al. (2024) & Generative AI & \\
\cite{Ostaszewski2021Reinforce}~Ostaszewski et al. (2021) & Reinforcement Learning& \cite{liu2023quantum}\\
\cite{wilson2021optimizing} Wilson et al. (2021) & Meta Learning & \cite{2019arXiv190705415V}\\
\hline
\end{tabular}
\end{table}

\section{AI for Device Control and Optimization}\label{sec:control_calibration} 
All approaches to building and operating quantum processors involve control, tuning and optimization of quantum devices. Control refers to actively modifying quantum states through inputs (e.g. microwave pulses) to perform desired operations. Tuning involves adjusting device parameters to target a specific operating regime, and optimization involves refining such parameters to maximize performance metrics like coherence times, operation speeds, and fidelity. The characterization of quantum devices requires probing their properties to inform control, tuning, and optimization decisions.

\begin{figure}[h]
\centering
 \includegraphics[width=\textwidth]{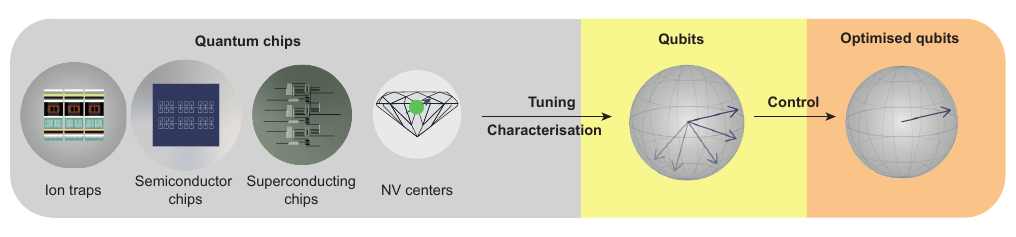}
\caption{Most quantum device architectures require specific tuning and control protocols to operate as qubits. Machine learning-based approaches allow us to automate and speed up such protocols, allowing for high-throughput characterization and optimization of quantum devices.}
\label{fig:Section4}
\end{figure}

In practice, the characterization, tuning, control and optimization of quantum devices are time-consuming processes, currently often requiring the dedicated work of a team of quantum physicists. The use of ML approaches for automating these processes is well motivated, since NNs and Bayesian optimization methods excel at inferring appropriate outputs from limited input data without employing costly modeling from first principles. A variety of ML methods have been used to characterize different types of quantum devices, automate tuning strategies, and optimize qubit control. The ability to automate a full tuning pipeline, including the encoding of the qubit and its optimization, is a fundamental requirement for most QC platforms (see Fig.~\ref{fig:Section4}). Most of the automation work has focused on tackling different stages of the tuning pipeline. In Table~\ref{table:control}, we summarize some key references in which ML-based algorithms for tuning, characterization and optimization of quantum devices have been demonstrated. 

Semiconductor quantum dot devices are a good case-study for the application of AI to tuning methods and control strategies, due to the large number of parameters that are still required to encode and operate spin qubits. Many ML methods have been explored to automate and optimize the operation of these devices. A variety of classifiers and NNs have been used to tune and identify charge transitions in large parameter spaces~\cite{
kalantre2019ml, nguyen2021deep, vanEsbroeck2020quantum} and detect Pauli Spin blockade~\cite{SchuffPSB2023} (a step often required for spin qubit initialization and readout). Automated strategies based on Bayesian optimization have proven robust for tuning quantum dot devices from scratch (i.e. tuning from a de-energized device to a double quantum dot configuration - often referred to as super coarse tuning)~\cite{moon2020machine, straaten2022all}. Bayesian methods have also been used for multiparameter cross-compensation~\cite{hickie2023automated}, and for quantum device tuning across different material systems~\cite{severin2021cross, Severin:2023crl}. 
Recently, the development of algorithms involving the interplay of Bayesian optimization, CNNs  and computer vision has allowed the demonstration of the first complete tuning of a single spin qubit~\cite{schuff2024fully} and the optimization of qubit Rabi speed and coherence time~\cite{carballido2024qubit}.
Bootstrapping techniques can also be employed with these models, to further reduce the amount of input data required~\cite{
wozniakowski2020boosting} and enable partial inference based on the findings of previous studies. 

Some qubit control techniques rely on characterizing the environment of a qubit. Real time learning of Hamiltonian parameters, enabled by fast adaptive Bayesian estimation has been employed for such characterizations. In turn this has enabled the optimization of qubit control - resulting in extended coherence times both in semiconductor~\cite{berritta2024realtime, berritta2024PRApp} and NV-centers spin qubits~\cite{scerri2020extending, ArshadPRapp2024}.

In superconducting quantum circuits, NNs can be used to relax the requirement of device characterization and control complexity, instead modeling the single-trajectory output of a qubit directly, with high accuracy~\cite{koolstra2022monitoring, Flurin2020}. In superconducting qubits NNs with deep and RL, have been used to demonstrate cavity state preparation~\cite{porotti2022deep}
, mid-circuit measurements~\cite{vora2024ml}, quantum control~\cite{metz2023self}, and qubit initialization~\cite{reuer2023}. Model-free RL, in which a learning agent is trained using only measurement outcomes, has also been used to demonstrate quantum control of a superconducting qubit~\cite{sivak2022model} and quantum error correction in a 3D superconducting cavity~\cite{sivak2023nat}.
In photonics platforms, ML-based methods have been implemented for control and characterization of both single and multiparameter systems~\cite{cimini2019calibration, cimini2021calibration}.

\begin{table}[h!]
\footnotesize
\begin{tabular}{p{2.5cm}p{4cm}p{1.5cm}p{3cm}}
\hline
\textbf{Platform} & \textbf{Application} & \textbf{Approach} & \textbf{Reference} \\
\hline
Trapped ions & Enhanced qubit readout & NN-classifier & \cite{seif2018machine} Seif et al. (2018) \\
BEC & Non-linear readout & NN & \cite{cao2022quantum} Cao et al. (2023) \\
Semiconductor Nanowire & Fully automatic tuning of a spin qubit & BO, CNN, CV & \cite{schuff2024fully} Schuff et al. (2024) \\
 & Qubit speed/coherence optimisation & BO, CNN, CV & \cite{carballido2024qubit} Carballido et al. (2024) \\
Semiconductor QDs 
& DQD tuning & DCNN & \cite{kalantre2019ml} Kalantre et al. (2019) \\
 & Qubit regime tuning & BO & \cite{teske2019ml} Teske et al. (2019) \\
 & DQD tuning and disorder characterization & BO & \cite{moon2020machine} Moon et al. (2020) \\
 & Identification of transport features & Deep RL & \cite{nguyen2021deep} Nguyen et al. (2021) \\
 & Fine tuning of transport features & UL & \cite{vanEsbroeck2020quantum} van Esbroeck et al. (2020) \\
 & Cross-platform tuning & BO-RF & \cite{severin2021cross} Saverin et al. (2024) \\
 & Spin readout identification & NN, CNN & \cite{SchuffPSB2023} Schuff et al. (2023) \\
 & All RF-DQD tuning & BO & \cite{straaten2022all} van Strateen et al. (2022) \\
 & DQD tuning with charge sensor compensation & BO & \cite{hickie2023automated} Hickye et al. (2023) \\
 & Characterisation of electrostatic disorder potential & BO, CNN & \cite{craig2024bridging} Craig et al. (2024) \\

 & Hamiltonian learning and real time qubit control & BO & \cite{berritta2024realtime} Berritta et al. (2024a) \\
 & Characterization of qubit fluctuations & BO & \cite{berritta2024PRApp} Berritta et al. (2024b) \\
NV-centers & Characterization of qubit fluctuations & BO & \cite{ArshadPRapp2024} Arshad et al. (2024) \\
& Hamiltonian learning and real time qubit control & BO & \cite{scerri2020extending} Scerri et al. (2020) \\
& Nuclear spin sensing with spin qubit & DL-NN & \cite{jung2021deep} Jung et al. (2021) \\
Superconducting qubit & Qubit characterization & NN & \cite{wozniakowski2020boosting} Wozniakowski et al. (2020) \\
 & Quantum dynamics reconstruction & RNN & \cite{Flurin2020} Flurin et al., (2020) \\
 & Cavity state preparation & DRL & \cite{porotti2022deep} Porotti et al. (2022) \\
 & Mid-circuit measurements & Multilayer-NN & \cite{vora2024ml} Vora et al. (2024) \\
 & Qubit initialization & RL-NN & \cite{reuer2023} Reuer et al. (2023) \\
 & Quantum control & Model-free RL & \cite{sivak2022model} Sivak et al. (2023) \\
3D superconducting cavity & Quantum error correction & Model-free RL & \cite{sivak2023nat} Sivak et al. (2023) \\
\hline
\end{tabular}

\caption{Key experimental demonstrations of ML-based algorithms for quantum device tuning, characterization and optimization.}
\label{table:control}
\end{table}

 \section{AI for Quantum Error Correction}\label{sec:qec}

Salable quantum error correction (QEC) is a critical prerequisite for FTQC yet extremely difficult to realize in practice. The following sections explore how AI may improve the demanding decoders needed to run QEC and help accelerate the discovery of more efficient QEC codes.

\subsection{Decoding}\label{sec:decoding}
QEC protocols involve making joint measurements on sets of qubits (syndrome qubits) and using these results to infer which physical qubits (data qubits) have most likely experienced errors. Locating errors allows them to be corrected or otherwise accounted for in the remainder of a computation. The inference step of this process is performed by a classical decoding algorithm, which has the difficult task of not only making a best-guess of error locations from limited information, but furthermore must do so with high enough speeds to prevent an insurmountable backlog of errors. Decoders are an important consideration for QEC codes, playing a part in determining the noise threshold below which a QEC code can suppress errors.

Decoders for QEC therefore face serious scalability challenges. Moreover they need to provide adaptability to address the diverse noise models applying to different qubit architectures. The strict time-frames in which decoding operations must complete are governed by qubit coherence times and further constrained by connection latency between the decoder-running classical hardware and QPU. Simulations have confirmed that as quantum systems scale, decoders struggle to meet the required low-latency thresholds~\cite{kurman2024controlrequirementsbenchmarksquantum}. This problem is exacerbated by increasing qubit count and the complexity of QEC codes. Furthermore, decoders often operate under the assumption of a simple noise model, typically a depolarizing channel. In practice, there are many far more complex real-world noise models - often including correlations. These considerations make it challenging for practical decoders to adapt effectively without significant performance loss.

A diverse array of AI techniques are being explored as tools to improve the efficiency, accuracy, and scalability of QEC decoding algorithms~\cite{torlai2017neural, chamberland2018deep, wagner2020symmetries, baireuther2018machine, sweke2020reinforcement, andreasson2019quantum, breuckmann2018scalable, ueno2022neo, wang2023transformer, bausch2023learning, chamberland2023techniques, lange2023data, mann2024mlmp, wang2023dgr}. The majority of this work targets the surface code and other topological codes with some of these decoders showing great promise for scalability. AI based decoders are also being explored for other quantum LDPC (qLDPC) codes.


AI-powered decoders were pioneered in work leveraging Boltzmann machines to decode various stabilizer codes with the ability to generalize between codes~\cite{torlai2017neural}. Later, a CNN approach was used to decode higher dimensional QEC codes, and showed strong evidence of scalability over other ML decoders, whilst also requiring less retraining as system sizes were varied~\cite{breuckmann2018scalable}. This CNN approach assigns local likelihoods of errors, producing a threshold of around 7.1\% for the 4D toric code under noiseless syndrome measurements. Subsequent work applied a CNN with binarization to realize higher efficiency~\cite{ueno2022neo}. More realistic approaches leveraged 3D convolutions with the temporal dimension (round of measurements). Such approaches take advantage of strategies like syndrome collapse and vertical cleanup to speedup a minimum-weight perfect matching (MWPM) decoding algorithm. This approach has been demonstrated  on a distance 17 surface code, running the protocol on local FPGA hardware~\cite{chamberland2023techniques}.

Similarly, a long short-term memory recurrent neural network (LSTM RNN) trained on experimentally accessible data outperforms a traditional MWPM surface code decoder by capturing correlations between bit-flip and phase-flip errors~\cite{baireuther2018machine}. This approach adapts to the physical system without requiring a noise model and maintains performance over multiple error correction cycles.  

\begin{figure}[h]
    \centering
    \includegraphics[width=\textwidth]{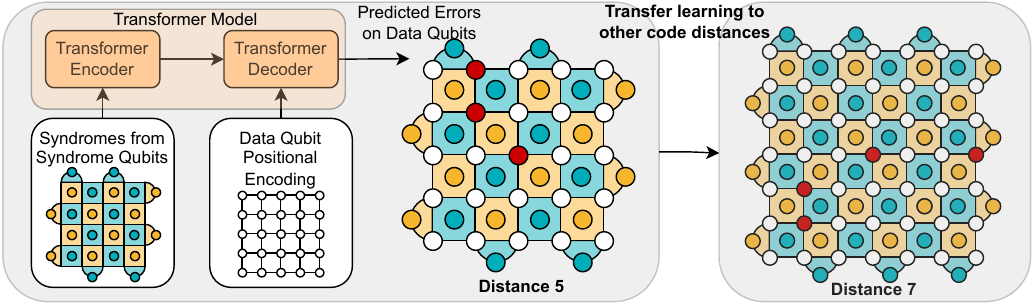}
    \caption{Transformer model for decoding quantum surface code. Figure adapted from~\cite{wang2023transformer}. Models trained on a small code distance (5 above) can transfer to larger distances (7 above) thanks to the variable input length of the transformer, cutting down on the training time.}
    \label{fig:transformer-qec}
\end{figure}

The first application of attention-based transformer models to decoding was presented in 2023, producing a logical error rate below baselines given by MWPM and Union Find decoders~\cite{wang2023transformer}. As shown in Figure~\ref{fig:transformer-qec}, a transformer encoder is used to embed  syndrome information and a transformer decoder is used to predict errors on each data qubit. Additional work added recurrent units to a transformer decoder and outperformed traditional approaches when trained on data from Google's Sycamore processor for code distances 3 and 5 \cite{bausch2023learning}.

Formulating decoding as a graph classification problem has also allowed graph neural networks (GNN's) to outperform traditional matching algorithms on the surface code, using simulated data and not requiring an explicit noise model \cite{lange2023data}. This GNN approach achieved competitive performance on experimental repetition code data. Though GNN training is onerous, inference is efficient and scales linearly, indicating great promise for fast, accurate, and noise model-free decoding in practical error correction. GNN decoders have also proven successful for qLDPC codes~\cite{mann2024mlmp}, outperforming traditional belief propagation and ordered-statistics decoding process. They have also demonstrated favorable transfer learning properties that allow decoding of high-distance codes with models trained on low-distance codes. 

More sophisticated approaches like RL have also shown great utility for decoding.  RL has demonstrated how an ML technique known as deep Q-learning can be applied to develop decoders that work with faulty syndrome measurements \cite{sweke2020reinforcement}. RL deep Q-networks have been used to decode toric code bit-flip errors with comparable performance to the MWPM algorithm for small error rates \cite{andreasson2019quantum}. Besides directly performing the decoding task, other work~\cite{wang2023dgr} leverages AI model to adjust the weights in decoder graph for drifted and correlated errors.

In summary, AI-based methods for QEC decoding offer significant potential to transform the field by addressing both scalability and adaptability challenges that conventional MWPM decoders struggle with. By leveraging advanced architectures—such as CNNs for spatial error likelihoods, transformers for scalable syndrome embedding, and GNNs for graph-based error localization—AI-powered decoders can achieve superior accuracy and faster inference without relying on different noise models. These methods adapt dynamically to complex noise environments, capturing correlations and variations more effectively than MWPM. They demonstrate strong evidence of outperforming traditional methods, providing a promising pathway toward efficient, scalable, and noise-resilient quantum error correction crucial for practical QC.
        
\subsection{Code Discovery}\label{sec:code_discovery}
The discovery of new QEC codes is crucial for advancing FTQC, especially as we strive to design codes that are more efficient, robust, and tailored to specific quantum hardware architectures. Traditionally, finding optimal QEC codes has been a labor-intensive process, relying on analytical approaches and domain expertise to explore the vast space of possible code structures. However, the complexity of quantum systems and the diversity of noise environments pose challenges that can be difficult to address with conventional methods. AI offers a promising alternative by automating the search for new QEC codes, leveraging its ability to identify patterns and optimize structures within high-dimensional spaces. ML models, such as neural networks and RL agents, can explore code design spaces far beyond human intuition, identifying novel error-correction schemes and optimizing parameters to meet desired performance criteria. This data-driven approach can potentially accelerate the discovery of QEC codes that offer improved error thresholds, scalability, and adaptability to real-world noise models, thus playing a critical role in the future of quantum computation.

RL has been used for discovering various QEC codes and can be enhanced for example to encode circuits tailored to specific hardware~\cite{olle2023discovery} in which RL agents have been shown to scale to 20 qubits and distance-5 codes. Noise-aware meta-agents can enable learning across various noise models, accelerating the discovery process. RL can also be used to optimize tensor network code geometries for discovering stabilizer codes~\cite{mauron2024optimization}. Using the projective simulation framework, the RL agent efficiently identifies optimal codes, including those with multiple logical qubits, outperforming random search methods. For instance, this RL agent finds an optimal code 10\% of the time after 1000 trials, compared to a theoretical 0.16\% from a random search - giving a 65-fold performance improvement. Other work~\cite{su2023discovery} has leveraged RL in the Quantum Lego~\cite{cao2022quantum} framework, focusing on maximizing code distance and minimizing logical error under biased noise. In this case an RL agent discovers improved code constructions, including an optimal $\left[\left[17,1,3\right]\right]$ code that better protects logical information. This approach allows for tailored code design without the need for explicit noise model characterization.

\section{AI for Postprocessing}\label{sec:AI_for_post}
Quantum applications commonly require a post-processing stage to extract meaningful results from quantum measurements and optimize the measurement process. The following sections explore how AI may improve efficient observable estimation, tomography, and readout measurements and how it can be applied to error mitigation techniques.

\subsection{Efficient Observable Estimation and Tomography}\label{sec:obs_est_tomo}
Estimating quantum observables is a key part of quantum computations, wherein quantum information is reduced into readable, classical information. Such measurement data comprises the entirety of what we may probe about a quantum system, but can be costly to obtain. Estimating an observable to some required accuracy entails combining samples from multiple measurements - with the number of required observables and samples scaling (possibly exponentially) in the system size under consideration \cite{anshu2024survey}. As such, ML and AI techniques have proven useful for reducing the quantity of data points needed to estimate a given observable, using the blackbox structure of AI models for more efficient inference.
        
Even in the case of simple models of a single observable, such as the reconstruction of a Rabi oscillation in a spin system, ML methods have been seen to improve measurement efficiency by reconstructing state populations with smaller sample sizes \cite{struck2021robust}.  These advantages persist for more complicated systems, such as the classification of multiqubit states, where NNs significantly enhance the accuracy of state classification, mitigating the effect of limitations such as qubit readout crosstalk \cite{seif2018machine,lienhard2022deep}. 

The measurement overhead of full quantum tomography \cite{d2003quantum} is even more computationally intensive than the estimation of quantum observables. Full state tomography (FST) is, in fact, impractical for all but the smallest quantum systems. Quantum state tomography (QST) has emerged as a more feasible solution to FST by focusing on alternative approaches like shadow tomography that address the exponential scaling cost of FST~\cite{torlai2018neural}. One of the most promising variations of QST is the use of NN-based approaches for state reconstruction. NNs can be applied to learn quantum states from limited measurement data, making them suitable for learning the state of large systems. A good example is ShadowGPT, which uses simulated shadow tomography data to train a GPT based model for predicting Hamiltonian ground state properties \cite{yao2024shadowgptlearningsolvequantum}. 



CNNs can be used for the reconstruction of high-fidelity quantum states with a fraction of the data that is traditionally required. For example, when applied to ground states of the transverse-field Ising model, a CNN-based tomography scheme achieves a tenfold reduction in observable estimation error compared to conventional maximum likelihood methods~\cite{schmale2022efficient}. This demonstrates enhanced accuracy with only polynomially scaling resources.

NN's can also be used in adaptive QST, through a neural adaptive quantum tomography (NAQT) method~\cite{quek2021adaptive}. This is an adaptive framework that applies RNNs to replace computationally intensive Bayesian updates. This use of RNN's dynamically optimizes the measurement strategy, allowing it to efficiently approximate the quantum state with far fewer resources. These efficiencies are critical for scaling QST to larger systems.

\subsection{Readout Measurements}
AI can significantly enhance the processing of qubit readout by improving measurement accuracy and minimizing the effects of noise and errors. For example, in neutral atom quantum computing systems, AI has been used to improve qubit state detection accuracy, where CNNs have outperformed traditional methods and achieved an up to 56\% reduction in readout errors for single-qubit measurements~\cite{phuttitarn2024enhanced}. 

In trapped ions and Bose Einstein condensate neural networks have been implemented to enhance qubit readout~\cite{seif2018machine} and realize non-linear readout schemes~\cite{CaoPRL2023}.

These AI-driven enhancements are particularly significant in the context of FTQC. High-fidelity readout is a key requirement for QEC because it relies on accurate measurements to detect and correct computational errors. Thus, AI helps to move closer to the precision levels necessary for FT operations.

    \subsection{Error Mitigation Techniques}

Quantum error mitigation (QEM) is a set of techniques that attempt to deal with noise in quantum systems without resorting to the full machinery of FTQC, which can include both NISQ and early FT hardware~\cite{caiQuantumErrorMitigation2023}. Rather than dealing with noise by introducing redundancy through extra qubits, QEM techniques mainly tackle errors through additional circuit runs. This approach relies on the fact that the goal of a quantum computation is often obtaining the noiseless observable expectation value, rather than the quantum state itself. By running additional circuits to probe the effects of various noise components on the target expectation value, QEM aims to reverse these inferred effects during post-processing. A large range of QEM techniques have been developed that utilize various aspects of the computation: knowledge of the gate noise (probabilistic error cancellation (PEC) and zero-noise extrapolation (ZNE))~\cite{liEfficientVariationalQuantum2017,temmeErrorMitigationShortDepth2017};  known purity and symmetry constraints on output states~\cite{bonet-monroigLowcostErrorMitigation2018,mcardleErrorMitigatedDigitalQuantum2019,hugginsVirtualDistillationQuantum2021,koczorExponentialErrorSuppression2021,liuVirtualChannelPurification2024}; or knowledge about the target problem (subspace expansion)~\cite{mccleanHybridQuantumclassicalHierarchy2017}. 

Many of these techniques contain hyperparameters which are usually obtained via device calibration. These hyperparameters can also be obtained by optimizing over a set of training circuits.
As is often the case with AI in other application areas, there are challenges in providing relevant training data. This is even more prominent in the quantum case since without a noiseless machine, it is impossible to obtain the ideal output of a general quantum circuit for the training examples. One approach to dealing with this is to train using classes of classically simulable circuits, such as Clifford circuits. One can prove that the mean square training loss over all Clifford circuits is actually equal to that over all possible unitary circuits since the set of Clifford circuits forms a unitary 2-design~\cite{strikisLearningBasedQuantumError2021}. Hence, by training on Clifford circuits alone, we can achieve optimal generalization power. However, in practice, it can be challenging to implement general Clifford circuits due to the number of gates required. Furthermore, we are often interested in the expectation value of a few target circuits rather than general unitaries. Hence, instead of using the full set of general Clifford gates, we can restrict training to circuits having similar structures and noise profiles to the target circuits of interest, such that their outputs are affected by the noise in a similar way. Such an approach has been successfully implemented in leading QEM techniques like PEC~\cite{strikisLearningBasedQuantumError2021} and ZNE~\cite{czarnikErrorMitigationClifford2021}. 

The direct application of NNs to QEM (in a similar fashion to ZNE) has also been explored~\cite{kimQuantumErrorMitigation2020}. Conventional ZNE constructs a model of how an observable's expectation value varies with noise, fitting parameters to this model by probing the expectation value at different circuit noise levels. Instead of explicitly constructing a model, the noisy expectation values of different circuit sizes can be directly input into a multi-layer perceptron, which then outputs the noiseless values of larger-size circuits. This allows one to train the NN on small classically simulable circuits, and then use it to make predictions for large, non-simulable, circuits. This approach has been further extended to other ML methods like random forest and GNNs and compared against conventional ZNE for practical problems in actual hardware for up to 100 qubits~\cite{liaoMachineLearningPractical2023}. It has been found that random forest models outperform other ML techniques, also outperforming (linear) ZNE in all cases. It is found that extrapolating random forest models to make predictions beyond circuit sizes within their training dataset leads to a distinct increase in errors. This problem has been approached by including larger training circuits with target expectation values supplied by hardware experiments applying conventional QEM~\cite{liaoMachineLearningPractical2023}. In this way, the ML model can mimic the behavior of the supplied QEM techniques, and lead to a lower sampling cost in experiments.

QEM can also be applied to NN tomography (see \cref{sec:obs_est_tomo}). Suppose one has found the circuit for preparing the ground state of some given Hamiltonian, but the output state is corrupted by noise. Approaches have been suggested~\cite{bennewitzNeuralErrorMitigation2022} for first characterizing the output noisy state using NN tomography and then further optimizing parameters in the NN to minimize the energy and mitigate the errors, in a similar spirit to subspace expansion~\cite{mccleanHybridQuantumclassicalHierarchy2017}. Note this requires using an autoregressive network so that the NN states can be sampled from directly in order to perform energy optimization. 

The various methods for applying AI to QEM described above have proven extremely fruitful but are still relatively limited in their scope compared to the large variety of conventional QEM techniques. However, taking advantage of other quantum circuit features, such as richer noise models and more detailed circuit structures, suggests the possibility of more novel future applications of ML in QEM.

\section{Outlook}\label{sec:outlook}

The research surveyed in this work demonstrates how AI has the potential to enable breakthroughs in virtually all aspects of the development and operation of quantum computers. AI techniques are not only useful in NISQ-era devices and applications, but will also play an essential role in the building of large-scale FT machines. The exploration of how AI can be of utility for quantum computing has only just begun, and by focusing more on these techniques the quantum community stands to see further breakthroughs in the challenges facing useful QC.   In this section we raise awareness of a number of areas of development that can catalyze improvement and further adoption of AI in QC.

\subsection{Accelerated Quantum Supercomputing Systems}

A prerequisite for researching and deploying AI models for quantum research is access to supercomputing resources. Increasingly sophisticated AI techniques require greater processing power to train, and classical computing capabilities will need to scale alongside developments in quantum hardware. This has driven international efforts to prioritize quantum research and in many cases, integrate physical quantum hardware within AI supercomputing infrastructures \cite{NQISRC2024,EuroHPC2024}.

\begin{figure}[h]
\centering
\includegraphics[width=0.8\textwidth]{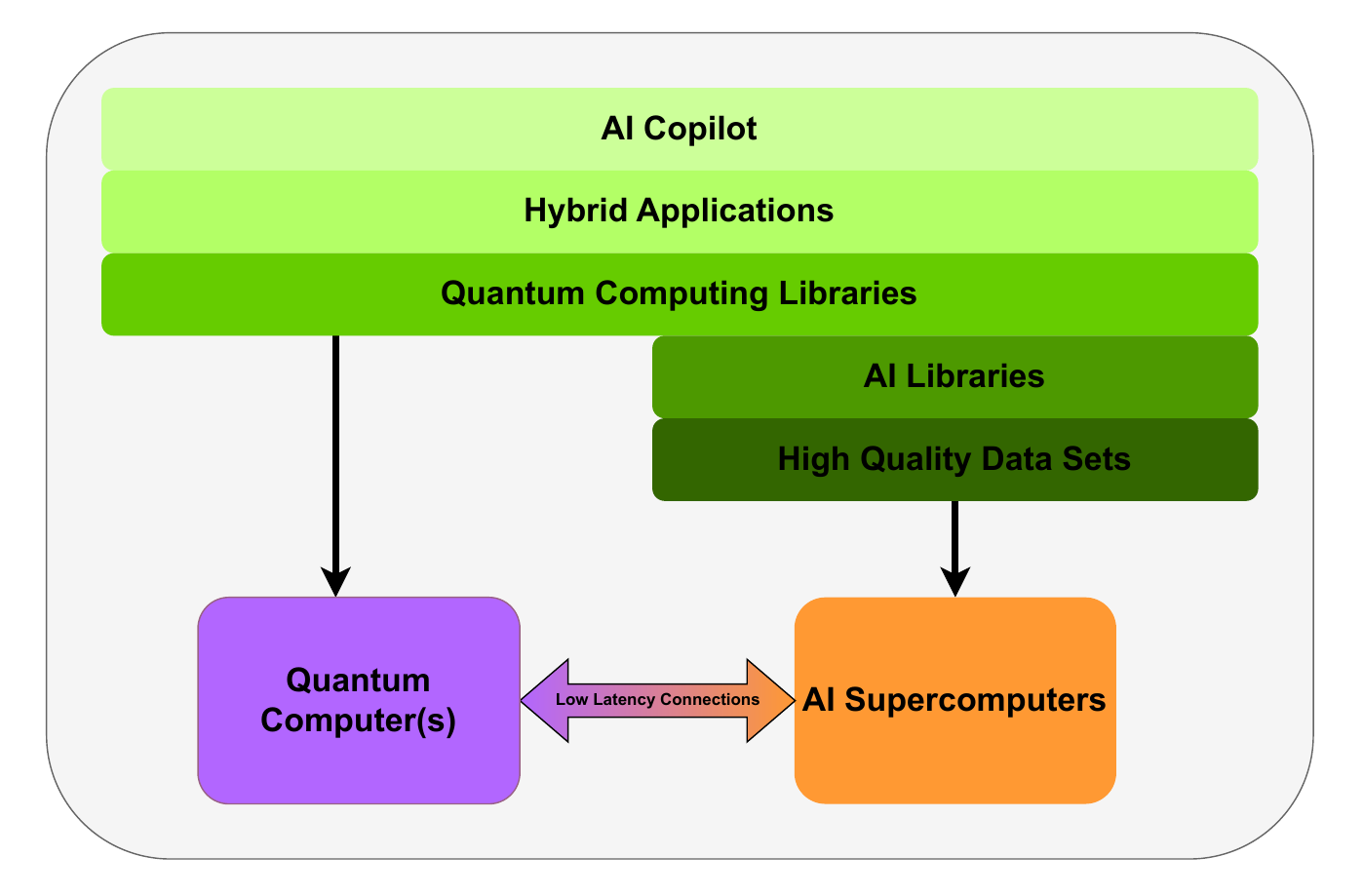}
\caption{Depiction of a development platform incorporating access to both QC and AI resources. Such a platform should be accessible to both domain scientists and quantum developers, and must orchestrate hybrid workflows leveraging both AI supercomputers and quantum processors.}
\label{fig:dev_platform}
\end{figure}

Such integration of quantum processors within AI supercomputers is widely accepted to be a necessary architecture for building large-scale, useful quantum computers. But doing so requires specialized software and in some cases, additional specialized hardware \cite{beck2024integrating, kurman2024controlrequirementsbenchmarksquantum}. 
Applying classical supercomputing to scaling challenges facing QC, such as QEC, requires extremely low-latency interconnects between collocated classical and quantum hardware. 

Development platforms for quantum-classical architectures must provide user-friendly hybrid programming workflows able to maximize performance across a heterogeneous compute architecture (see Fig.\ref{fig:dev_platform}). To ensure adoption, such a platform should support popular scientific computing and AI libraries as well as those specialized libraries required for domain specific applications and quantum devices control concerns. Given the broad and interdisciplinary set of users for such a platform, AI copilots will likely play a particularly important role in reducing the barrier to entry for domain experts unfamiliar with quantum application development.

\subsection{Simulating High Quality Data Sets}\label{sec:simulation}

Many applications of AI models for quantum computing must be trained on large, high quality data sets. This often entails experimental data from quantum systems, which 
is costly to obtain given the limited availability and capabilities of current quantum processors. There are efforts to democratize and reuse existing quantum data~\cite{placidi2023mnisqlargescalequantumcircuit, zwolak2024data}. But though this encourages collaboration and transparency, such efforts are unfortunately  unlikely to keep up with the increasing demand for training data.

This shortfall can be addressed by synthetic training data, obtained through simulation. Simulating quantum systems can come at exponential computational cost, but is a necessity while quantum hardware is still limited in scale, quality, and availability. Conventional supercomputing resources can be used to simulate quantum processors through techniques such as  quantum dynamics, digital density matrix, quantum trajectories (SV and TN) \cite{Koopman_1931, BenioffSV1, BenioffSV2, Feynman_1982, breuer2002theory, vidal_2003}, and large scale (Clifford, stabilizer, Pauli paths) \cite{gottesman1998, aaronsonclifford2, bravyi2016, Aharonov_2023, gonzálezgarcía2024}. GPU-acceleration of devices toy models can also be used to generate large training datasets with hundreds of thousands of samples~\cite{QArrayStraaten, QArrayStraaten_release}. 
As listed, these techniques can simulate systems of increasing scale, though at the cost of accuracy. Powered by accelerated computing they can generate large and diverse data sets necessary capable of fueling the applications covered in this review \cite{cuQuantumSDK}.

Simulation heuristics can improve the quality of synthetic data generated by simulations, and the efficiency with which it is created. ADAPT (Automatic Differentiation Adapted Quantum) methods, for example, iteratively construct quantum circuits based on the gradient of a cost function with respect to the gate chosen from an operator pool, which tends to provide resource-efficient circuits, which can be used to train other models. The methods mentioned in Sec.~\ref{sec:compact_circuits} provide similar benefits.

AI can also be used to simulate quantum systems directly and has driven substantial scientific progress not only in the field of QC, but in condensed matter, quantum chemistry, and similar fields. Neural quantum states (NQS) are among the most widespread of such simulation strategies \cite{carrasquilla2017machine,carleo2017solving,lange2024architectures, yang2024NQS}. NQS are large classical NN's that can be sampled like a quantum system to generate data. Unlike data-driven applications however, NQS can also be optimized {\it variationally}, with knowledge of a quantum system's Hamiltonian or Linbladian. For many applications, NQS have the potential to provide a more compact and scalable representation than other many-body techniques, such as tensor networks, but the benefits are still under investigation \cite{bukov2021learning}.

Several additional AI models have been developed specifically to solve time-dependent quantum dynamics. For example, the Heisenberg equations of motion can be modeled via deep NNs \cite{han2021tomography,mohseni2022deep}. More recently, Fourier Neural Operators (FNOs) \cite{shah2024fourier} have been deployed to learn the time-evolution of quantum spin systems. Of particular interest is the ability of FNOs to extrapolate to predictions spanning longer times than are seen in training, potentially extending quantum evolution past the capacities of modern devices or tractable tensor network depths. Additionally, gradient-based optimization protocols and Bayesian inference combined with differentiable master equation solvers have been proved useful to compute steady state solutions and time evolutions of open quantum systems~\cite{craig2024differentiable}.

\subsection{Increased Multidisciplinary Collaboration}

Availability of computational resources, development platforms, and quality data sets are all foundational for enabling AIs application to QC research. But another key gating factor is collaboration between AI and QC experts. It is likely that many of the most cutting edge AI techniques have not yet had their greatest impact on developments in QC, and many areas remain where deep collaboration may yield new breakthroughs. Here we highlight several areas in AI research we believe hold great potential for future exploration.


Diffusion models (introduced in Sec.~\ref{sec:AIMethodsSurvey}) have proven incredibly impactful in other application areas, but have so far only been applied to unitary synthesis~\cite{ho2020denoising} for quantum computing (see Sec.~\ref{sec:unitary_synthesis}). There is also the opportunity to apply recent training methodologies to problems in the development of quantum computing. For example, Advanced RL techniques have been proposed to solve real-world problems (See \cite{silver2017mastering,fawzi2022discovering} as examples). Whilst generative flow networks \cite{bengio2021flow} combine RL and generative models in an effort to incorporate reasoning in ML-based exploration. Both of these techniques are potentially useful in QC, where RL is already being applied. 

Perhaps one of the more ambitious uses of AI for quantum, and a prime opportunity for collaboration, is the design of new quantum algorithms. Quantum algorithm design is an extraordinarily difficult task, and could rightly be regarded as a grand challenge in science. All currently known quantum applications rely on a small set of algorithmic primitives, which has not been significantly expanded for many decades. One future strategy is the use of AI to generate novel quantum algorithms. Such a process could begin by defining a scientific problem and then working backward to generate the quantum circuits needed to provide a solution according to some figure of merit. Optimistically, one could hope that beyond an application-specific algorithm, such an approach might also lead researchers to entirely new algorithmic primitives and thus unlock whole new classes of quantum algorithms. This kind of AI-assisted approach to algorithm discovery would represent an evolution of the AI-driven circuit synthesis methods described in Sec.~\ref{sec:circuit_synthesis} and other alternative approaches \cite{xiao2021stochastic, zhu2023artificial, bang2014strategy}. 

It may also be the case that incorporating AI into the algorithm design process could lead to the generation of more fundamentally hybrid quantum algorithms, that draw more optimally on both quantum and AI-based algorithmic primitives. Continued collaboration between the AI and quantum computing communities could easily also result in the development of new models developed specifically for QC applications. The development of quantum-specific foundational models may see the greatest AI-enabled breakthroughs in QC yet.

\section{Conclusion}
 QC is experiencing an explosion of utility from AI. The research surveyed in this review demonstrates that AI can play a role in everything from designing qubits, preparing efficient quantum algorithms, controlling and calibrating the device, correcting errors in realtime, and interpreting the output from QC. Most importantly, each aspect of QC needs to scale, and AI might be the only tool with the ability to both solve these problems effectively and do so efficiently at scale. AI has only begun to benefit QC and it is likely that AI will play an increasingly critical role into the realization of useful QC applications and FTQC.

\backmatter
\section{Acknowledgments}
AAG thanks Anders G. Frøseth for his generous support and acknowledges the generous support of Natural Resources Canada and the Canada 150 Research Chairs program. S.C.B and Z.C. acknowledge the EPSRC projects EP/T001062/1, EP/W032635/1, and EP/Y004655/1. N.A. acknowledges support from the European Research Council (grant agreement 948932) and the Royal Society (URF-R1-191150). N.A. and F.F. also acknowledge support from the European Union and UK Research \& Innovation (Quantum Flagship project ASPECTS, Grant Agreement No.~101080167). Views and opinions expressed are however those of the authors only and do not necessarily reflect those of the European Union, Research Executive Agency or UKRI. Neither the European Union nor UKRI can be held responsible for them. RGM acknowledges that Research at Perimeter Institute is supported in part by the Government of Canada through the Department of Innovation, Science and Economic Development Canada and by the Province of Ontario through the Ministry of Economic Development, Job Creation and Trade.

\bibliographystyle{naturemag}
\bibliography{bibliography}

\end{document}